\documentclass[aps,twocolumn,prc,showpacs]{revtex4}
\usepackage{epsfig}
\begin{document}
\title{Neutron-neutron correlation in the halo dissociation of light exotic nuclei}
\author{M. T.  Yamashita$^a$, T. Frederico$^b$, and Lauro Tomio$^c$}
\affiliation{$^a$Unidade Diferenciada de Itapeva, Universidade Estadual Paulista,
18409-010 Itapeva, Brasil, }
\affiliation{$^b$Departamento de F\'\i sica, Instituto
Tecnol\'ogico de Aeron\'autica, Centro T\'ecnico Aeroespacial,
12228-900 S\~ao Jos\'e dos Campos, Brasil,}
\affiliation{$^c$Instituto de F\'\i sica Te\'orica, Universidade Estadual Paulista,
01405-900 S\~{a}o Paulo, Brasil}
\date{\today}
\begin{abstract}
We present model results for the two-halo-neutron correlation
functions, $C_{nn}$, for the dissociation process of light exotic nuclei
modelled as two neutrons and a core. A minimum is predicted for
$C_{nn}$ as a function of the relative momentum of the two neutrons,
$p_{nn}$, due to the coherence of the neutrons in the halo and
final state interaction.
Studying the systems $^{14}$Be, $^{11}$Li and $^{6}$He within this
model, we show that the numerical asymptotic limit, $C_{nn}\to 1$,
occurs only for $p_{nn}\gtrsim$ 400 MeV$/c$, while
such limit is reached for much lower values of $p_{nn}$
in an independent particle model as the one
used in the analysis of recent experimental data.
Our model is consistent with data once the experimental correlation
function is appropriately normalized.
\end{abstract}
\maketitle
The discovery of radioactive exotic weakly bound nuclei, rich in
neutrons or protons, beyond the drip line have brought a lot of interest
in the nuclear structure and reactions of these unstable nuclei.
The traditional nuclear models are unable to describe the long-range
correlations between the nucleons of the halo. The large size of the
halo has stimulated several experimental and theoretical studies devoted
to clarify the new aspects of the structure, stability and reaction of these
nuclei, including astrophysical applications (see, for example,
\cite{zhu-pr93,nie-pr01,jensen-rmp04}).

The effect of the large spatial extension of the neutron halo, was
probed in a recent fusion experiment of $^{6}$He with $^{238}U$
target~\cite{ali-nat04}. It was observed a large reaction
cross-section due to a direct $2n$ ($n$ represents a neutron)
transfer rather than an enhancement of the complete fusion cross-section.
In view of the large size of the halo
($r^{rms}_{nn}=5.9\pm1.2$~fm~\cite{mar-plb00}, $r^{rms}_{nn}$ is the
$nn$ root-mean-square radius) which is comparable to the size of $^{238}U$
itself, it is most likely that a correlated neutron-neutron ($nn$) is
transferred to $^{238}U$ while the $^4$He
is still far from the fusion barrier, favoring the $2n$ transfer process.

A novel view of the complete fusion reaction for large halo nuclei
also emerges~\cite{hinde-nat04}: it is most probable that the core
approaches the fusion barrier of an excited target nucleus that
has already absorbed the halo. This interpretation is in line
with the concept of an absorptive many-body potential written in
the relative coordinates of the core, two neutrons and the
target~\cite{hus-np04}.

It is evident the need of a deeper insight into the key aspects
of the structure and dynamics of the halo in Borromean three-body
systems (where all the subsystems are unbound), like $^6$He.
In this respect, using intensity interferometry applied with a new
iterative technique, Marqu\'es et al.~\cite{mar-plb00,mar-prc01}
recently probed the spatial configuration of two-neutron halo systems
and estimated the mean-square $nn$ distances as well, considering
the dissociation of $^{6}$He, $^{11}$Li and $^{14}$Be in the field
of a heavy nucleus target.  The spatial configuration of $^{11}$Li
was also studied by Petrascu et al.\cite{petra-np04,petra-prc04}.
The $nn$ correlation function, $C_{nn}$, is extracted as a function of
the relative momentum between the neutrons, $p_{nn}$.
As the absolute normalization of the correlation function is an
important piece of physical information, one should consider a consistent
model for that. We note that, for the asymptotic normalization, in the
fit to data, it has been considered a model where the halo neutrons are
assumed to be independent~\cite{LL}. In our understanding, one should
assume them as being emitted by a coherence source.

In the present communication we report our results for the $C_{nn}$ of
Borromean three-body systems $n-n-A$, where $A$ is the core mass
number. We consider $^6$He, $^{11}$Li and
$^{14}$Be within a description of two neutrons and a core forming
Borromean systems. We show how to circumvent the major difficulty of
the relationship between the initial and final states due to the
distorting effects of the reaction: we consider the $nn$ final state
interaction (FSI) and the three-body structure of the Borromean
system. The $nn$ FSI is shown to play a crucial role in distorting
the relative motion of the neutrons and it is a source of
interference effects which leads to an unexpected minimum of the
$C_{nn}$, as a function of $p_{nn}$, pushing the asymptotic behavior
(where $C_{nn}=1$) to larger momentum.
By comparing our results with the available experimental ones~\cite{mar-plb00,mar-prc01,petra-np04}, we found that it is
reasonable to expect a different normalization for the data, as will
be shown.

We use a three-body model in the limit of zero-range interaction,
which retains the essential physics of the weakly bound and large
two-neutron halo systems~\cite{nie-pr01,jensen-rmp04}.
The interaction singularity is tamed in a renormalized zero-range
model~\cite{amo-prc92,ad95,amo-prc97,yama-np04} which is appropriate
to study weakly bound three-body systems. The model is parameterized
by minimal number of physical inputs, which are directly related to
known observables: the two-neutron separation energy, $S(2n)=-E_{nnc}$,
the $nn$ and neutron-core ($n-A$) $s-$wave scattering lengths
(or the corresponding virtual or bound state energies).

The neutrons of the halo have a large probability to be found
outside the interaction range. Therefore the low-energy properties
of these halo neutrons are, to a large extend, model independent as
long as few physical input scales are fixed. The model provides a
good insight into the three-body structure of halo nuclei, even
considering some of its limitations. It is restricted to
$s-$wave two-body interactions, with small energies for the bound or
virtual states.
Even in this case, the three-body wave function for the valence neutrons
presents configuration mixing due to angular momentum recoupling
[see also Refs.~\cite{amo-prc97,Zin95}].
We also note that, all the interaction effects, such as higher partial waves
in the interaction and/or Pauli blocking effect are, to some extend,
included in the model, as long as the experimental two- an
three-body energies are supplied.
In this perspective, it is of interest to extend the three-body
continuum calculation of $C_{nn}$, done in \cite{dan-prc04}
for $^6$He, to analyze the actual data.

The zero-range interaction itself does not have a physical scale,
then any calculated observable only depends on the two and
three-body energy scales. The dependence on the three-body scale
is a consequence of the Thomas collapse~\cite{th35} of the
three-body system in the limit of a zero-range interaction which
demands a three-body scale to supply the system with the physical
information of the three-particle short distance configurations.
Therefore, any dimensionless observable is a function of the ratios
between the input energies, the mass ratio of the core and the
neutron $(A)$ and as well as on the nature of the subsystems,
bound or virtual~\cite{ad95}. The functional form of each
observable written in terms of dimensionless ratios is given by a
universal scaling function valid for short ranged interactions or
large systems~\cite{yama-pra03}. The model is more general and can
describe phenomena in atomic and molecular physics such that other
large three-body systems with different structures can also be
treated and stability and size studies performed (see e.g.
Ref.~\cite{jensen-rmp04,bra04}).

We show that the asymptotic limit $C_{nn}\to 1$ is reached
in our model at much higher values of $p_{nn}$ than the
ones found in previous data
analysis~\cite{mar-plb00,mar-prc01,petra-np04,petra-prc04}.
Due to coherence of the neutrons in the halo
and final state interaction, $C_{nn}$ goes smoothly to
the asymptotic limit only after displaying a minimum, as it
will be shown.

Particle distributions in the halo were  calculated
in Refs.~\cite{yama-np04,yama-pra03}, where the three-body Faddeev
equations for the renormalized zero-range two-body interactions
were solved. We obtain $C_{nn}$ by using the corresponding
three-body wave function with the inclusion of the $nn$
final-state interaction (FSI)~\cite{gelbke}.
We will show our results in case of the halo
nuclei $^{6}$He, $^{11}$Li and $^{14}$Be.

For the $n-n-A$ three-body system, $C_{nn}$ is given by
\begin{eqnarray}
&&C_{nn}\equiv C(\vec{p}_{nn})=\frac
{\int d^3q_{nn}|\Phi(\vec{q}_{nn},\vec{p}_{nn})|^2}
{\int d^3q_{nn}\rho(\vec{q}_{n A})\rho(\vec{q}_{n^\prime A})},\\
&&\vec{q}_{nA}\equiv\vec{p}_{nn}-\frac{\vec{q}_{nn}}2 \;\;\;{\rm and}\;\;\;
\vec{q}_{n^\prime A}\equiv-\vec{p}_{nn}-\frac{\vec{q}_{nn}}2,\nonumber
\end{eqnarray}
where the one-body density is
\begin{eqnarray}
\rho(\vec{q}_{nA})=\int d^3q_{n^\prime A}
\left|
\Phi\left(-\vec{q}_{nA}-\vec{q}_{n^\prime A},\frac{\vec{q}_{nA}-
\vec{q}_{n^\prime A}}{2}\right)
\right|^2.
\end{eqnarray}
$\Phi\equiv\Phi(\vec{q}_{nn},\vec{p}_{nn})$ is the corresponding
breakup amplitude of three-body wave function including the FSI
between the neutrons. $\vec{q}_{nn}$ is the relative momentum between the
core $A$  and the
center-of-mass of the $nn$ subsystem; and $\vec{p}_{nn}$ the
relative momentum between the neutrons.

The FSI is introduced directly in the inner product
$\Phi\equiv\langle \vec{q}_{nn};\vec{p}_{nn}\,^{(-)}|
\Psi\rangle$, where the ket $|\vec{p}_{nn}\,^{(-)}\rangle$ refers
to the $nn$ scattered wave given by the Lippmann-Schwinger
equation. The correlation function, calculated with the
distorted-wave amplitude, assumes a sudden breakup of the halo
as the main reaction mechanism. The halo is considered as a
coherent source of neutrons, differently from the framework of
Ref.\cite{LL}. In our picture, the slow halo motion decouples
from the fast motion of the core in the field of the target.
The distorted wave amplitude $\Phi$ is given by
\begin{eqnarray}
\Phi=\Psi(\vec{q}_{nn},\vec{p}_{nn})+\frac{1/(2\pi^2)}{\sqrt{E_{nn}}-ip_{nn}}
\int d^3p
\frac{\Psi(\vec{q}_{nn},\vec{p})}{p_{nn}^2-p^2+i\epsilon},
\label{dwampl}
\end{eqnarray}
where $\Psi$ is the three-body wave function~\cite{yama-np04}.
$E_{nn}$ is the $nn$ virtual state energy taken as 0.143 MeV.

In the framework of Lednicky-Lyuboshits~\cite{LL}, in order to 
 obtain the correlation function $C_{nn}$, the probability density 
 (in configuration space) of the neutron-neutron scattering state is 
 multiplied by the probability density of the relative motion of the 
 halo neutrons in the three-body wave function. 
 Such model for the correlation function was developed from 
 astrophysics, and the possibility to apply it to particle and nuclear 
 physics is valid if one assumes that the particles emitted by the source
 are independent.
 We believe this is not the case, when considering the neutrons of the
 halo of a nuclei. So, in our approach, when the distorted-wave amplitude 
 of Eq.~(\ref{dwampl}) is considered, $|\Phi|^2$ contains off-diagonal matrix      elements of the two-body densities.

\begin{figure}[thbp!]
\centerline{\epsfig{figure=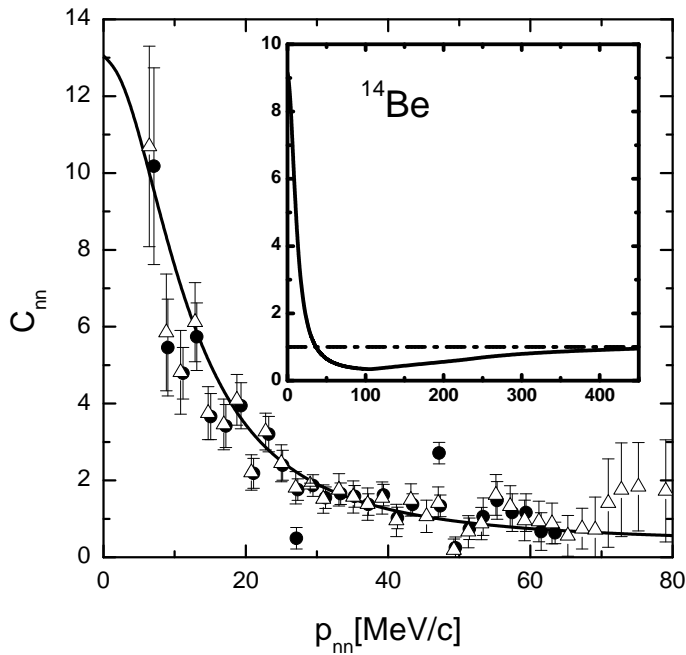,width=9cm}} \caption{
Two-neutron correlation for the halo of $^{14}$Be, as a function
of the relative $nn$ momentum, $p_{nn}$.
The solid curve gives the model results for $S_{2n}=$ 1.337 MeV, and
$E_{nA}$=0.2 MeV. When compared with data, the model result is
multiplied by 1.425.
Experimental data are from \cite{mar-plb00} (open triangles) and
\cite{mar-prc01} (full circles). } \label{fig1}
\vskip 0.2cm
\end{figure}

\begin{figure}[thbp!]
\centerline{\epsfig{figure=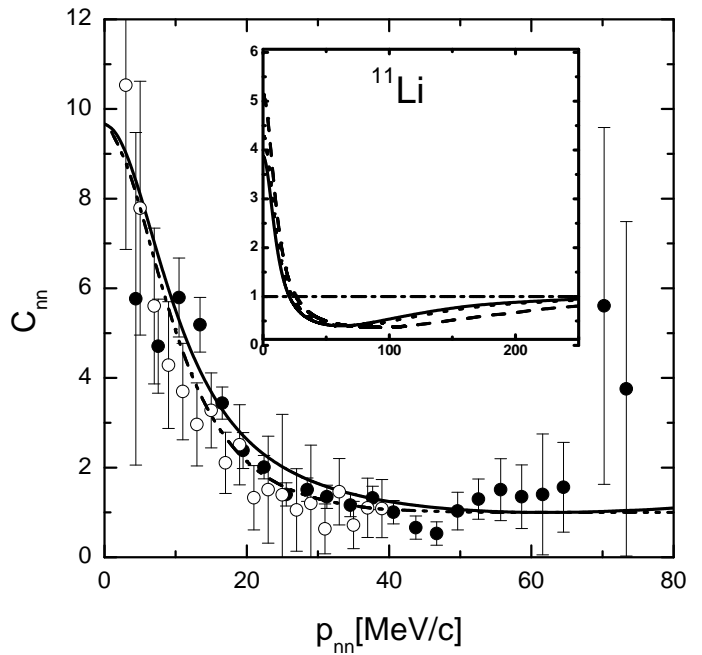,width=9cm}}
\caption{
Two-neutron correlation for the halo of $^{11}$Li, as a function
of the relative $nn$ momentum, $p_{nn}$.
The model results (inset) are given for three cases:
$S_{2n}=$ 0.29 MeV and $E_{nA}$=0.05 MeV (solid line);
$S_{2n}=$ 0.37 MeV and $E_{nA}$=0.8 MeV (dashed line); and,
$S_{2n}=$ 0.37 MeV and $E_{nA}$=0.05 MeV (dotted line).
In the main body of the figure, the solid curve
(when $r^{rms}_{nn}=8.5$fm \cite{yama-np04}) presents
the corresponding curve of the inset multiplied by 2.5;
the dot-dot-dashed curve, the model presented in
\cite{petra-np04} with $r^{rms}_{nn}=8.3\;$fm.
The experimental data are from \cite{mar-plb00}
(full circles) and \cite{petra-np04} (empty circles).
} \label{fig2}
\vskip 0.2cm
\end{figure}

\begin{figure}[thbp!]
\centerline{\epsfig{figure=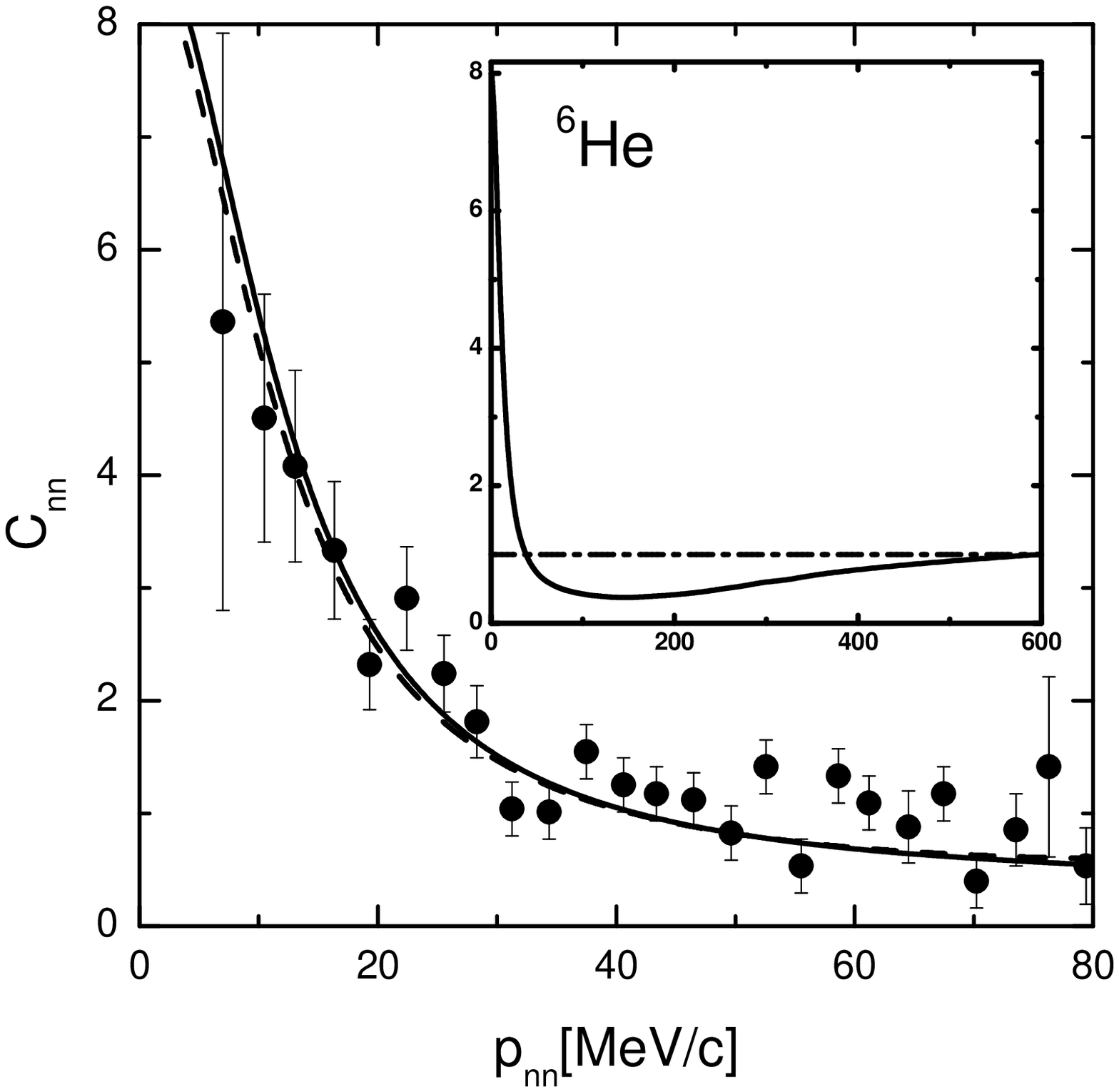,width=9cm}} 
\vskip 0.2cm
\caption{
Two-neutron correlation for the halo of $^6$He, as a function of
the relative $nn$ momentum, $p_{nn}$. The solid curve gives the
model results  for $S_{2n}=$ 0.973 MeV and $E_{nA}$=4~MeV; 
and, the dashed  curve, for $S_{2n}=$ 0.973 MeV and $E_{nA}$=0.
When compared with data (extracted from \cite{mar-plb00}), 
the model results (in the main body) are conveniently normalized 
as explained in the text.}
\label{fig3}
\vskip 0.2cm
\end{figure}

The results of our calculations for $C_{nn}$ of the systems
$^{14}$Be, $^{11}$Li and $^6$He are respectively shown in
Figs.~\ref{fig1} to \ref{fig3}. They are shown as functions
of $p_{nn}$, and compared with experimental available data~\cite{mar-plb00,mar-prc01,petra-np04}.

In the present Faddeev renormalized zero-range model, as expected,
$C_{nn}$ goes asymptotically to unity, as a function of $p_{nn}$.
Such limit is shown clearly in all the three selected systems that
we have analyzed. The results that include the asymptotic limit
are shown in the insets of the figures. As we observe, for
example, in the insets of Figs.~\ref{fig1} to ~\ref{fig3}, the
theoretical $p_{nn}$ asymptotic limits for the systems $^{14}$Be,
$^{11}$Li, and $^6$He are, respectively, $\gtrsim$ 400  MeV$/c$,
$\gtrsim$ 250 MeV$/c$, and $\gtrsim$ 500 MeV$/c$.
Besides the fact of the
qualitative similar behavior, when comparing the results of the
insets with the corresponding experimental results, we note a
clear scale discrepancy. The striking
observation is that the interference effect produced by the
inclusion of FSI originates a minimum for $C_{nn}\approx$ 0.35,
pushing the asymptotic limit to much larger values of $p_{nn}$
than the ones considered in the asymptotic normalization of the
experimental data.
Our model calls for a different
normalization of the experimental $C_{nn}$ obtained in Refs.~\cite{mar-plb00,mar-prc01,petra-np04}.
We believe that the qualitative picture presented by our model,
evidencing a minimum of $C_{nn}$, will survive in a more realistic
three-body approach.

In Figs.~\ref{fig1} to \ref{fig3}, we observe a quite good
agreement between theory and data when an appropriate normalization factor
is included. Particularly, for $^{14}$Be, the data are well reproduced by the
model with $S_{2n}$=1.337 MeV~\cite{audi95} and with $E_{nA}=$0.2~MeV~\cite{tho00}
(where $E_{nA}$is the virtual state energy of $n-^{12}$Be), when our $C_{nn}$ is
multiplied by 1.425, as shown in Fig.~\ref{fig1}
(or, $C_{nn}\approx 0.70\; C_{nn}^{exp}$,
where $C_{nn}^{exp}$ is the model fit to experimental data).

We have studied the model sensitivity to the two and three-body
binding energies considering the case of $^{11}$Li. We chose this
nucleus because it has the smallest $2n$ separation energy
$S_{2n}$ among the nuclei we are analyzing. The model results are
shown by the three plots presented in the inset of
Fig.~\ref{fig2}. For a fixed $s-$wave $^{10}$Li two-body virtual
state energy, $E_{nA}=$0.05 MeV~\cite{Zin95}, we vary $S_{2n}$
from 0.29 MeV~\cite{tan96} (solid line) to 0.37
MeV~\cite{lunney04} (dotted line). They differ only near the
origin. Next, to observe the sensitivity to the $s-$wave $^{10}$Li
two-body virtual state energy, we also plot a curve for
$E_{nA}=$0.8 MeV~\cite{wil75} and $S_{2n}=$ 0.37 MeV (dashed
line). Increasing the virtual state energy, the ``plateau" near
the minimum is enlarged and, for $p_{nn}<$ 60 MeV$/c$, $C_{nn}$ is
enhanced and even strongly near zero. 

We should note that, as we increase the 
virtual state energy (reduce the absolute value of $n-^9$Li
scattering length) for a fixed $S_{2n}$ in $^{11}$Li,
we are shrinking the three-body system (see discussion and Table 1
of Ref.~\cite{yama-np04}), implying that $C_{nn}$ develops a longer 
tail in momentum space. 
In the inset of Fig.~\ref{fig2}, this effect 
is clear and shown by the variation of the $n-^9$Li virtual state energy from 
0.05 to 0.8 MeV, while $S_{2n}$ is kept fixed at 0.37 MeV. 
The increase of $S_{2n}$ from 0.29 to 0.37 MeV, with  $E_{nA}$fixed
at 0.05 MeV, also implies in reducing the size of the three-body
system. This is shown in the inset of Fig.~\ref{fig2} by comparing the solid
with dotted lines; the variation is small but consistent with the
small variation of $S_{2n}$.
We expect that such typical behavior is general. With respect to the correct
asymptotic behavior, one can observe our solid line given in the
inset figure. We need to multiply it by a factor 2.5 in order to
fit the experimental data, as shown in the main body of the
figure. So, in this case we have $C_{nn}\approx 0.40\;
C_{nn}^{exp}$. We also compare our calculation with another model,
given in \cite{petra-np04}, which is close to the one presented in
\cite{mar-plb00} in the framework of \cite{LL}. We observe that
the different behavior between our and other models, essential for
the absolute normalization of the data, will appear at higher
momentum values.

Next, to study the case of $^6$He, it's worthwhile to mention that the
$s-$wave $n-^4$He scattering has a positive scattering length and effective
range~\cite{nhe73,neutronscl}, which by extrapolation of the
standard effective range expansion produces an unphysical $^5$He
bound state. Similar situation is found in the quartet
neutron-deuteron scattering where the large positive scattering
length and an effective range produces, by naive extrapolation, a
bound state near the threshold. The effective interaction in the
case of the $n-d$ system is repulsive and no quartet trinucleon
exists, similarly for the $n-^4$He where the $s-$wave interaction
should be effectively repulsive to avoid an unphysical bound
state. However, in our model the sign of the $n-$core scattering
length is determinant of the occurrence of the bound state.

The experimental $n-^4$He scattering length is 3.26(3)~fm~\cite{neutronscl},
which by naive extrapolation of the effective range in our zero-range model
produces an unphysical bound state. Therefore, we arbitrarily made a
calculation of $C_{nn}$ with a zero energy $s-$wave $^5$He and
the experimental value of $S_{2n}$ in $^6$He. Surprisingly, we found a reasonable
agreement with the data, as shown in Fig.~\ref{fig3}. In this case, our
results for $C_{nn}$ need to be multiplied by 1.48 to fit the data,
as shown by the dashed line in the main part of the figure
($C_{nn}\approx 0.68\; C_{nn}^{exp}$).

Observe that the interaction for the $n-^4$He should be weaker than the one we
are using; and, as the interaction becomes weaker while the
three-body energy is kept constant, the three-body system tends to
be more compact~\cite{yama-np04} implying in an increase of the
typical three-body momentum scale. To mimic this effect, we decrease
the attraction of the $n-^4$He $s-$wave interaction, allowing a
virtual $s-$wave state (unphysical), using as a typical magnitude
for the scattering length the value of 3 fm. Still in this case,
as we have multiplied our $C_{nn}$ by 1.12,
we obtain a good fit to data, as shown by the solid line of Fig.3.
So, leaving the normalization free, one cannot distinguish between
different weakly attractive $s-$wave interactions in the $^5$He, if
the $S_{2n}$ and the $nn$ virtual state energy are fixed.

Alternatively, one may think in using the zero-range $T-$matrix and
project out the bound-state, using the experimental value of the
$n-^4$He scattering length. This may effectively weaken the
interaction, which could in principle produce a similar effect as
we have discussed.

In conclusion, we presented a theoretical study of neutron-neutron
correlation for Borromean three-body weakly-bound systems,
considering three cases where experimental data are available. As
the present analysis of halo dissociation in a renormalized
zero-range three-body framework contains the main physical scales
of weakly bound Borromean systems, we believe that our results for
low $nn$ relative energies (below 40 MeV) are to a large extend
universal; i.e., other short-ranged potential models, with the
same low energy scales, will produce similar results. Considering
that in our model there are no free parameters, as the inputs are
just the physical scales (two and three body observables), and
$C_{nn}(p_{nn})\to 1$, when $p_{nn}\to\infty$, one can use it to
get the normalization of the data. 
 Our zero-range calculations suggest that
the relevant effect of the inclusion of final-state interactions
with the halo considered as a coherent source of neutrons is the
presence of a minimum in $C_{nn}$ at an intermediate region of
$p_{nn}$ (see Figs.~\ref{fig1}-\ref{fig3}). 
It will be important to verify this conclusion, as well as possible
deviations, within a three-body model with more realistic 
two-body interactions. 
Finally, we observe a good agreement between theory and
experimental results, provided that the data fits, $C_{nn}^{exp}$,
are appropriately normalized. In view of our findings, it will be
very interesting to improve the experiments~\cite{mar-plb00,
mar-prc01, petra-np04, petra-prc04}, particularly at higher
momenta, in order to characterize the existence of the minimum in
$C_{nn}$. 

We would like to thank Prof. M. Petrascu for details on results shown in
\cite{petra-np04} and for a clarifying discussion.
For partial support, we also thank Funda\c c\~ao de
Amparo \`a Pesquisa do Estado de S\~ao Paulo and Conselho Nacional de Desenvolvimento Cient\'\i fico e Tecnol\'ogico.

\end{document}